\documentclass{article}

\PassOptionsToPackage{numbers, compress}{natbib}
\usepackage[preprint]{neurips_2023}

\usepackage[utf8]{inputenc} 
\usepackage[T1]{fontenc}    
\usepackage{hyperref}       
\usepackage{url}            
\usepackage{booktabs}       
\usepackage{amsfonts}       
\usepackage{amsmath}        
\usepackage{nicefrac}       
\usepackage{microtype}      
\usepackage{xcolor}         
\usepackage{float}   
\usepackage{graphicx}
\usepackage{booktabs}
\usepackage{placeins}
\usepackage[font=small]{caption}
\usepackage{caption}
\usepackage{hyperref}

\usepackage{newfloat}
\usepackage{caption}
\usepackage{bm}
\usepackage{stmaryrd}

\DeclareFloatingEnvironment[fileext=frm,placement={!htbp},name=Functionality]{Functionality}
\DeclareFloatingEnvironment[fileext=frm,placement={!htbp},name=Protocol]{Protocol}
\DeclareFloatingEnvironment[fileext=frm,placement=h,name=Protocol]{Protocol2}
\usepackage[framemethod=TikZ]{mdframed}
\usepackage{lipsum}

\newdimen\linenumbersep

\newcommand{\linenumber}[1]{%
  \linenumbersep 4pt%
  \advance\linenumbersep\mdflength{innerleftmargin}%
  \advance\linenumbersep\mdflength{innerlinewidth}%
  \advance\linenumbersep\mdflength{middlelinewidth}%
  \advance\linenumbersep\mdflength{outerlinewidth}%
  \advance\linenumbersep\mdflength{linewidth}%
  \makebox[0pt][r]{{\rmfamily\tiny#1}\hspace*{\linenumbersep}}}

\usepackage{dashrule}

\mdfdefinestyle{FunctionalityFrame}{%
    linecolor=black,
    outerlinewidth=0.5pt,
    font=\normalsize,
    roundcorner=2pt,
    innertopmargin=2pt,
    innerbottommargin=2pt,
    innerrightmargin=4pt,
    innerleftmargin=3pt,
    nobreak=true
    }

\mdfdefinestyle{ProtocolFrame}{%
    linecolor=black,
    outerlinewidth=0.5pt,
    font=\normalsize,
    roundcorner=1pt,
    innertopmargin=2pt,
    innerbottommargin=2pt,
    innerrightmargin=4pt,
    innerleftmargin=3pt,
    nobreak=true
    }

\usepackage{algpseudocode}

\usepackage{tabularx}
\usepackage{array,etoolbox}
\preto\tabular{\setcounter{magicrownumbers}{0}}
\newcounter{magicrownumbers}

\usepackage{booktabs}
\usepackage{amssymb}
\usepackage{pifont}
\usepackage[flushleft]{threeparttable}

\newcounter{protocol}


\usepackage[sets, operators]{cryptocode}

\definecolor{custompurple}{RGB}{102, 0, 153}
\definecolor{customred}{RGB}{139, 0, 0}
\definecolor{customblue}{RGB}{0, 0, 139}
\definecolor{customred2}{RGB}{200, 0, 0}

\usepackage{ulem}
\usepackage{multirow} 
\usepackage{booktabs} 
\usepackage{setspace}
\usepackage{graphicx}
\usepackage{url}
\usepackage{multicol}
\usepackage{amsmath}
\usepackage{tikz}
\usepackage{xspace}
\usepackage{enumitem}
\usepackage{adjustbox}
\usepackage{graphicx}

\definecolor{backgroundcolor}{HTML}{fff7e6}

\def\BibTeX{{\rm B\kern-.05em{\sc i\kern-.025em b}\kern-.08em
    T\kern-.1667em\lower.7ex\hbox{E}\kern-.125emX}}

\def\BibTeX{{\rm B\kern-.05em{\sc i\kern-.025em b}\kern-.08em
    T\kern-.1667em\lower.7ex\hbox{E}\kern-.125emX}}

\usepackage[linesnumbered,ruled,vlined]{algorithm2e}
\SetKwFor{RepTimes}{repeat}{times}{end}
\SetAlgoHangIndent{0pt}
\let\oldnl\nl
\newcommand{\nonl}{\renewcommand{\nl}{\let\nl\oldnl}}

\title{Agentic Privacy-Preserving Machine Learning\thanks{A position paper. Under active development.}}

\author{%
Mengyu Zhang, \quad Zhuotao Liu\thanks{zhuotaoliu@tsinghua.edu.cn}, \quad Jingwen Huang, \quad Xuanqi Liu
\\
InspiringGroup @ Tsinghua University
}

\begin{document}

\maketitle
\begin{abstract}
Privacy-preserving machine learning (PPML) is critical to ensure data privacy in AI. 
Over the past few years, the community has proposed a wide range provably-secure PPML schemes that relies on various cryptography primitives. 
However, when it comes to large language models (LLMs) with billions of parameters, the efficiency of PPML is everything but acceptable. For instance, the state-of-the-art solution for confidential LLM inference
represents at least 10,000-fold slower performance compared to plaintext inference. The performance gap is even larger when the context length increases. 

In this position paper, we propose a novel framework named Agentic-PPML to make PPML in LLMs practical. Our key insight is to 
employ a general-purpose LLM for intent understanding and delegate cryptographically-secure inference to specialized models trained on vertical domains. 
By modularly separating language intent parsing—which typically involves little or no sensitive information—from privacy-critical computation, Agentic-PPML completely eliminates the need for the LLMs to process the encrypted prompts, enabling practical deployment of privacy-preserving LLM-centric services.  

\end{abstract}
\section{Introduction}
Recent advances in large language models (LLMs) have fundamentally transformed how users interact with computational tools and data services. By integrating structured coordination protocols such as the Model Context Protocol (MCP) \cite{anthropic2024mcp}, LLMs can now function as intelligent orchestrators, capable of invoking external models or APIs on behalf of users. This enables seamless access to specialized functionality without requiring users to interact directly with underlying systems.
A key advantage of this framework is improved usability: rather than manually selecting or configuring a remote model provider, users can simply express their intent in natural language. The LLM interprets the request, identifies the appropriate MCP server, and formulates a structured query to facilitate the interaction automatically.

In practice, however, most MCP servers are hosted by third parties, raising serious concerns regarding possible leakage of sensitive user data and proprietary model content. Ensuring privacy in such settings requires careful integration of cryptographic mechanisms into the MCP workflow—a task that presents both technical and architectural challenges.
On the technical side, we aim to support efficient computation while preserving privacy. In general, there exists a fundamental tension between computational efficiency and strong privacy guarantees, as cryptographic protocols often incur substantial overhead.
On the architectural side, the challenge lies in embedding such protocols into the modular, stateless interaction pattern defined by MCP, where the LLM orchestrator must coordinate encrypted data flow between mutually distrustful parties without breaking abstraction boundaries or exposing sensitive content. This requires new abstractions that allow cryptographic protocols to be invoked as tools—just like any other API—while respecting the security model of all participants.


To address the above challenges, we propose Agentic-PPML, a novel framework for privacy-preserving inference with large language models (LLMs). The framework leverages a general-purpose LLM to perform intent understanding and delegates secure inference to domain-specific encrypted models via the Model Context Protocol (MCP). This design leverages the complementary strengths of both model types: while LLMs excel at general natural language understanding across diverse scenarios, traditional models (like CNNs) trained on specific domains are often more effective for specialized tasks such as structured reasoning or domain-specific data processing \cite{lau2018dataset}. By combining them through a secure and modular architecture, Agentic-PPML enables accurate and privacy-preserving inference without compromising either user data or model confidentiality.

Our design inherits the modular structure of the MCP, in which the user interacts with a LLM through natural language. The LLM is responsible for parsing the user's query and orchestrating the interaction with the appropriate MCP server, while the actual model computation is delegated to the MCP server. 
We observe that the user’s natural language query typically does not contain privacy-sensitive information. To balance privacy and efficiency, our system applies encryption exclusively to the neural network inference executed by the MCP server. Building on this foundation, we design a high-performance privacy-preserving inference protocol that combines two classes of cryptographic primitives in a hybrid manner: two-party secure computation techniques are used to support non-linear operations such as activation functions and comparisons, while homomorphic encryption (HE) is employed to efficiently perform linear computations like matrix multiplications.
This hybrid design allows our framework to preserve strong privacy guarantees while achieving practical inference efficiency on a wide range of neural network architectures.

A key advantage of our framework lies in its ability to decouple general-purpose language understanding from privacy-sensitive model inference. Unlike approaches that require deploying large LLMs under expensive privacy-preserving constraints (e.g., encrypted or federated LLM inference), our design allows the LLM to operate entirely on non-sensitive natural language queries in plaintext, without ever accessing user data or model internals. All privacy-critical content is handled exclusively by specialized MCP backends, where compact, task-specific models operate over encrypted inputs under rigorous cryptographic protection. This modular delegation allows us to replace costly private LLM inference with lightweight, privacy-preserving inference using vertically specialized models tailored to the task. To the best of our knowledge, this is the first system design that explicitly separates private inference from language coordination, enabling practical deployment of privacy-preserving AI services with improved scalability, efficiency, and usability.

\section{Overview}

\begin{figure}
    \centering
    \includegraphics[width=1\linewidth,trim={0cm 2cm 0cm 5cm}]{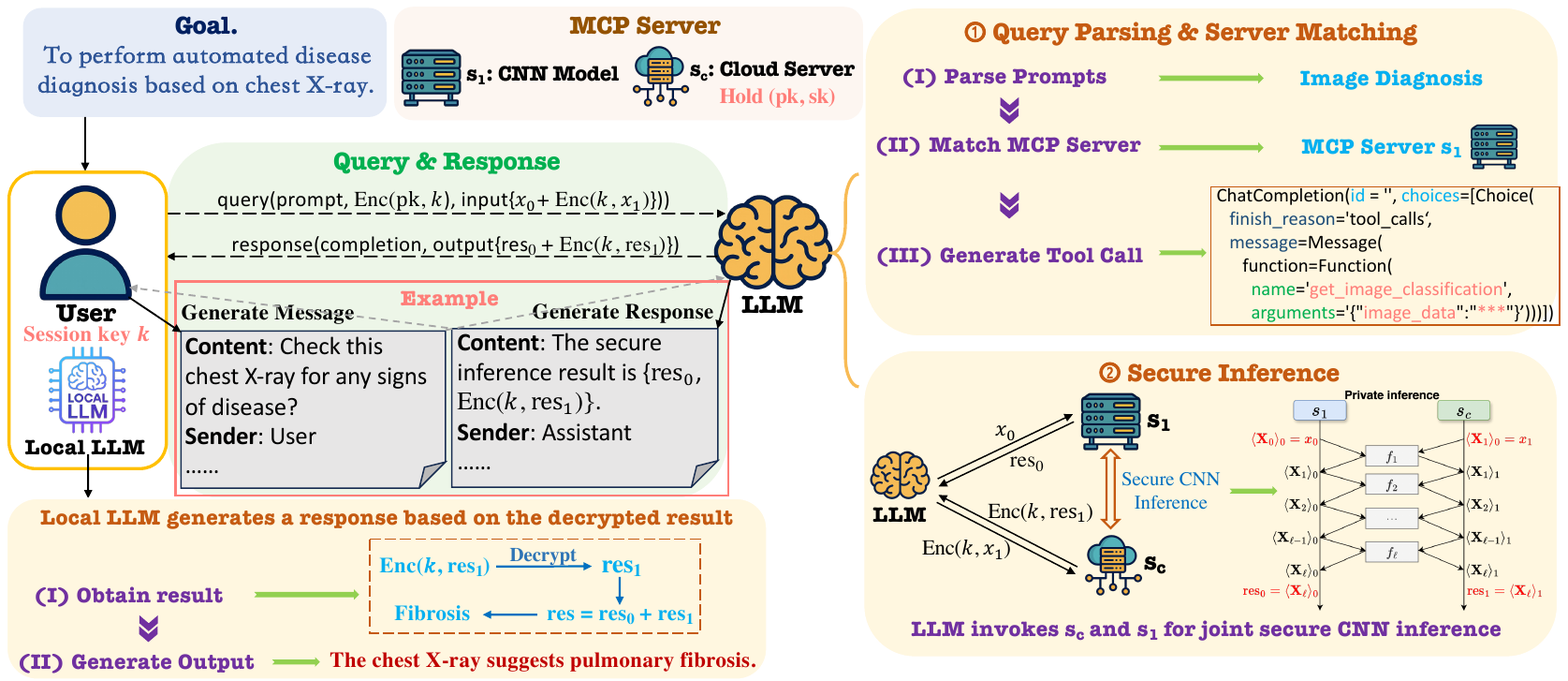}
    \caption{The overall system architecture.}
    \label{fig:enter-label}
\end{figure}

The overall system architecture is illustrated in Figure \ref{fig:enter-label}.
Our framework involves three primary roles: the user, the LLM, and the MCP servers.
In this setup, a user initiates a query and holds a piece of  private data that is intended to be processed by a remote MCP server. This private data remains encrypted throughout the protocol to preserve data confidentiality.
The LLM functions as a natural language interface and coordination agent. Upon receiving the user's query, the LLM understands the intent of the query and determines which MCP server is capable of handling the task. 
The MCP server, which hosts the model relevant to the query, engages in an MPC-based secure inference protocol with a cloud computing platform through inter-server coordination. In this setup, the cloud provider offers computational resources via its own MCP server, enabling collaborative inference while preserving both input and model confidentiality.

The entire computation protocol can be divided into four phases.
In the first phase, the user constructs a natural language query and prepares the corresponding inference input. For example, the user may submit a question such as “Could you help identify any issues in my chest scan?” to the LLM. Simultaneously, the user generates a one-time session key $k$, typically a symmetric key, which will be used to encrypt private input data and later decrypt the final inference result. To enable secure delivery of this key, the session key $k$ is encrypted under the cloud MCP server's public key and transmitted—via the LLM—to the cloud MCP server.
Next, the user prepares the inference input 
$x$ in a secret-shared form by splitting it into two additive shares, $x_0$ and $x_1$, such that $x = x_0 + x_1$. The share $x_0$ remains in plaintext, while the share $x_1$ is encrypted under the session key $k$. Both shares—$x_0$ and the encrypted $x_1$—are transmitted to the LLM, which coordinates the subsequent steps of the protocol.

In the second phase, the LLM parses the user query and selects the appropriate MCP server for downstream secure inference. Upon receiving the request, the LLM analyzes the query and identifies the underlying task, for example, chest radiograph diagnosis, which requires inference from convolutional neural networks (CNN). It then searches the registry of available MCP servers and selects a server $s_1$ that supports privacy-preserving CNN inference.
Since server $s_1$ performs secure CNN inference using a multi-party computation (MPC) protocol, the LLM constructs a response containing two function calls: one targeting the selected MCP server $s_1$, and the other targeting the cloud MCP server $s_c$. These calls jointly initiate the secure inference procedure over encrypted input, without revealing sensitive information to either party.

In the third phase, the selected MCP servers jointly perform privacy-preserving CNN inference. The LLM dispatches the two input shares by sending the plaintext share $x_0$ to the model-hosting MCP server $s_1$, and the encrypted share 
$x_1$ to the cloud MCP server $s_c$. Upon receiving the ciphertext, $s_c$ decrypts $x_1$ using the session key $k$ to recover the original share.
At this point, both $s_1$ (holding $x_0$) and $s_c$ (holding $x_1$) collaboratively execute a privacy-preserving machine learning inference protocol secured by applied cryptography. The protocol ensures that neither party learns the complete input $x$, nor do they gain access to each other's intermediate computations or the model parameters.
The inference output $\mathrm{res}$ is secret-shared into two additive components, $\mathrm{res}_0$ and $\mathrm{res}_1$, such that $\mathrm{res} = \mathrm{res}_0 + \mathrm{res}_1$. $s_1$ directly returns $\mathrm{res}_0$ to the LLM, while $s_c$ encrypts $\mathrm{res}_1$ using 
$k$ and transmits the ciphertext to the LLM.

In the fourth phase, the user securely reconstructs the inference result and generates a natural language response using a locally deployed LLM. Upon receiving the two result shares, the user decrypts the encrypted share $\mathrm{res}_1$ using the session key $k$, and combines it with the plaintext share $\mathrm{res}_0$ to recover the final inference output in plaintext.
The user then forwards both the original natural language query and the recovered inference result to a local LLM running on their own device. This local LLM synthesizes the final response, providing a user-facing output in natural language while ensuring that sensitive data never leaves the user’s trusted computing environment.

\section{Background}
In this section, we introduce the core cryptographic techniques that underpin the fundamental building blocks of our secure inference protocol. 

\subsection{Homomorphic Encryption}
Leveled Fully Homomorphic Encryption (LFHE) \cite{cryptoeprint:2012/144} is a public-key encryption scheme characterized by its ability to support the homomorphic evaluation of arithmetic circuits up to a specified depth $D$ on encrypted inputs. The scheme relies on the hardness of problems like the Learning With Errors (LWE) problem and is composed of four main algorithms,
each fulfilling a specific role:
\begin{itemize}
    \item $\mathrm{KeyGen}(1^\lambda) \rightarrow(\mathrm{pk}, \mathrm{sk})$: This algorithm takes a security parameter $\lambda$ and generates a corresponding public key pk and secret key sk.
    \item $\mathrm{Enc}(\mathrm{pk}, m) \rightarrow c$: The encryption process involves taking the public key pk and a message $m$, then outputting a ciphertext $c$. The message is assumed to belong to the space $\mathbb{Z}_p$ for some prime $p$.
    \item $\mathrm{Dec}(\mathrm{sk}, c) \rightarrow m$: This algorithm uses the secret key sk to recover the original message $m$ from the ciphertext $c$.
    \item $\mathrm{Eval}(\mathrm{pk}, c_1, c_2, f) \rightarrow c^{\prime}$: Given the public key pk , two ciphertexts $c_1$ and $c_2$ (encrypting messages $m_1$ and $m_2$, respectively), and a depth $D$ arithmetic circuit $f$, this algorithm produces a new ciphertext $c^{\prime}$.
\end{itemize}

\subsection{Additive Secret Sharing}
Additive secret sharing is a cryptographic method that divides a value $x \in \mathbb{Z}_p$, where $p$ is a prime, into multiple shares such that the original value can be reconstructed by summing the shares. In the 2-of-2 additive secret sharing scheme, $x$ is represented as a pair $\left(\langle x\rangle_1,\langle x\rangle_2\right)=(x-r, r) \in \mathbb{Z}_p^2$, where $r$ is a uniformly random element from $\mathbb{Z}_p$. The scheme guarantees that $x=\langle x\rangle_1+$ $\langle x\rangle_2 \bmod p$.
This construction is perfectly hiding, as each individual share $\langle x\rangle_1$ or $\langle x\rangle_2$ reveals no information about $x$. The randomness of $r$ ensures that the distribution of each share is uniform over $\mathbb{Z}_p$, making it impossible for an adversary with access to only one share to infer any information about $x$.

\subsection{Oblivious Transfer}
Oblivious Transfer (OT) is a fundamental cryptographic primitive that plays a central role in secure two-party computation. In a standard 1-out-of-2 OT protocol \cite{rabin1981ot}, a sender holds two input messages, while a receiver provides a choice bit and learns only the selected message, without revealing their choice to the sender. At the same time, the sender remains oblivious as to which message was retrieved.

With recent advances in OT extension and protocol optimization, the cost of performing large numbers of OTs has been significantly reduced \cite{iknp2003otextension, bcg19}, enabling practical deployment in real-world privacy-preserving applications. In this work, we adopt the RTT23 protocol \cite{10.1007/978-3-031-38551-3_19}—an optimized OT extension scheme based on BCG+19 \cite{bcg19}—as the foundation for our two-party secure computation. We implement this using the CrypTFlow2 \cite{cryptflow2} and SIRNN \cite{sirnn} toolchains, which provide high-performance and well-audited OT-based protocols for secure inference.

\section{Secure Machine Learning Inference}
In this section, we present the design of our secure machine learning inference protocol in a modular fashion. Rather than describing the protocol as a monolithic black box, we decompose it into a set of fundamental building blocks—each corresponding to a standard component commonly found in machine learning models, such as linear layers, activation functions, and normalization steps.

\subsection{Fully Connected Layer}
Matrix multiplication in fully connected (FC) layers is one of the core operations in neural network inference. A standard FC layer computes the transformation:
\begin{equation}
    Y=W \cdot X+b,
\end{equation}
where $W$ is the weight matrix and $b$ is the bias vector.
To support privacy-preserving inference, the input $X$ is secret-shared between the two computing parties, where $X_0$ is held by the MCP server and $X_1$ is held by the model owner. Then, the FC layer computation can be rewritten as:
\begin{equation}
    W \cdot X = W \cdot X_0 + W \cdot X_1, 
\end{equation}
where the second term can be computed locally by the model owner, and the first term requires a secure matrix multiplication protocol between the two parties. To perform the matrix product securely, we adopt a packing-based homomorphic encryption scheme. The idea is to encode matrix entries as polynomials using a multidimensional packing strategy and then apply homomorphic operations to compute partial results. After obtaining the encrypted matrix product, the parties jointly perform secure bias addition and complete the forward pass of the FC layer. 

\subsection{Convolution Layer}
Convolution is a fundamental operation in convolutional neural networks, commonly used to extract spatial features from structured inputs such as images. In this section, we describe how to securely evaluate a 2D convolutional layer within our MPC-based framework.
Let $X$ denote the input tensor and $W$ the convolution kernel. For a convolution layer, the forward computation is defined as: 
\begin{equation}
    Y=\mathrm{Conv2d}(W,X)+b,
\end{equation}
where $b$ is a bias vector broadcast and added across all output spatial positions. As in the fully connected case, $X$ is additively secret-shared between the MCP server and the model owner. As a result, the convolution operation is decomposed as: 
\begin{equation}
    \mathrm{Conv2d}(W,X) = \mathrm{Conv2d}(W,X_0) + \mathrm{Conv2d}(W,X_1),
\end{equation}
where the second term can be computed locally by the model owner, and the first term requires a secure protocol to compute the convolution over the secret-shared input. 
To enable efficient privacy-preserving convolution, we adopt a polynomial packing technique similar to the one used for matrix multiplication. In particular, both the weight tensor $W$ and the input tensor $X_0$ are flattened and encoded into multivariate polynomials according to a predefined layout that preserves convolutional structure. Once encoded, the homomorphic convolution can be evaluated efficiently using ring-based homomorphic encryption. 

\subsection{Batch Normalization Layer}

Batch normalization (BatchNorm) is widely used in deep neural networks to accelerate training convergence and improve generalization. In our privacy-preserving setting, we support secure inference through a carefully designed protocol that handles the affine transformation in BatchNorm securely under secret sharing and homomorphic encryption.

Let the input tensor be $X \in \mathbb{Z}_p^{B \times C \times D}$, where $B$ is the batch size, $C$ is the number of channels, and $D$ is the number of spatial dimensions per channel. The batch normalization layer applies a channel-wise affine transformation to the input:
\begin{equation}
    \hat{Y}_{b,c,d} = \frac{X_{b,c,d} - \mu_c}{\sigma_c} \cdot \widetilde{W}_c + \widetilde{b}_c,
\quad b \in [B],\ c \in [C],\ d \in [D]
\end{equation}
where $\mu_c$ and $\sigma_c$ are the pre-computed per-channel mean and standard deviation, and $\widetilde{W}_c$, $\widetilde{b}_c$ are learned affine parameters for channel $c$.
To implement this in the privacy-preserving setting, we pre-absorb the normalization into constants using:
\begin{equation}
W_c \gets  \frac{\widetilde{W}_c}{\sigma_c}, \quad
b_c \gets \widetilde{b}_c - \frac{\mu_c \cdot \widetilde{W}_c}{\sigma_c},
\end{equation}
so that the BatchNorm transformation becomes a linear operation:
\begin{equation}
\hat{Y}_{b,c,d} = W_c \cdot X_{b,c,d} + b_c.
\end{equation}

\subsection{ReLU Activation Function}
The ReLU (Rectified Linear Unit) function is one of the most commonly used non-linear activation functions in modern neural networks. Its standard definition is:
\begin{equation}
    y = \mathrm{ReLU}(x) = \mathrm{max}(x, 0),
\end{equation}
where $x$, $y$ denote the input and output values, respectively.
To realize the $\mathrm{max}(\cdot)$ function in the privacy-preserving setting, both parties first perform a secure comparison to determine whether the input is non-negative. This is done by invoking:
\begin{equation}
    \langle d \rangle \gets \mathrm{Positive}(\langle x \rangle),
\end{equation}
where $\langle d \rangle$ is a boolean secret share and $\langle x \rangle$ is the secret sharing form of $x$.
Finally, the $\mathrm{ReLU}$ output is computed via a secure multiplication:
\begin{equation}
    \langle y \rangle \gets \langle d \rangle \cdot \langle x \rangle.
\end{equation}

\subsection{Average Pooling Layer}
Pooling layers are used in convolutional neural networks to aggregate local information and reduce the spatial resolution of feature maps. In our protocol, we support average pooling as the forward operation, as it is more amenable to secure computation than max pooling.
Let the input tensor be $\tilde{X} \in \mathbb{Z}_p^{B \times C \times H \times W}$, 
where $B$ is the batch size, $C$ is the number of channels, and $H \times W$ is the spatial resolution of each feature map. 
Given a pooling kernel of size $H_k \times W_k$, the output tensor 
$\tilde{Y} \in \mathbb{Z}_p^{B \times C \times \left[\frac{H}{H_k}\right] \times \left[\frac{W}{W_k}\right]}$ 
is computed by taking the average over each non-overlapping region:
\begin{equation}
\label{equ: pooling}
\tilde{Y}_{b,c,i,j} = \frac{1}{H_k W_k} \sum_{u = iH_k}^{(i+1)H_k - 1} \sum_{v = jW_k}^{(j+1)W_k - 1} \tilde{X}_{b,c,u,v},
\quad b \in [B],\ c \in [C],\ i \in \left[\frac{H}{H_k}\right],\ j \in \left[\frac{W}{W_k}\right].
\end{equation}
The summation operations in Equation \ref{equ: pooling} can be computed locally by both parties using additive secret sharing, while the division operation requires invoking the Divide protocol based on oblivious transfer.
Thus, the complete privacy-preserving forward computation of average pooling is expressed as:
\begin{equation}
\langle Y_{b,c,i,j} \rangle \leftarrow \text{Divide} \left(
\sum_{u = iH_k}^{(i+1)H_k - 1} \sum_{v = jW_k}^{(j+1)W_k - 1} \langle X_{b,c,u,v} \rangle,\ H_k \times W_k
\right),
\end{equation}
where both the input and output are represented as secret shares.

\subsection{Discussion}

The modular building blocks described above—such as linear layers, convolutions, activations, normalization, and pooling—constitute the core components found in a wide range of modern neural network architectures. By designing each module to operate securely under a unified MPC framework, our system is not restricted to a fixed inference pipeline. Instead, it supports arbitrary compositions of these modules, enabling secure evaluation of diverse neural networks including convolutional networks, residual networks, and multi-branch architectures.

Importantly, the modularity of our design allows for continuous extension. As new neural components and architectural patterns emerge—such as attention mechanisms, normalization variants, or non-standard activation functions—they can be incorporated as additional secure subprotocols without modifying the overall framework. This extensibility makes our system well-suited for evolving machine learning workloads, and lays the foundation for a general-purpose privacy-preserving inference engine capable of supporting an expanding family of models.

\section{Experiment}

\subsection{Experimental Setup}
All experiments are conducted on a machine equipped with an Intel(R) Xeon(R) Gold 6348 CPU and an NVIDIA A100 GPU. The GPU supports hardware-accelerated CUDA-based cryptographic computation primitives required by our framework. We use CUDA version 12.1, as our implementation depends on features introduced in CUDA 11.7 and above to support 64-bit integer arithmetic in CUDA core functions.
The system runs Ubuntu 20.04 LTS. The codebase is compiled with GCC 11.4 for C++ components, Python 3.8.10 for high-level orchestration scripts, and Rust 1.77.0 stable for performance-critical secure computation modules.
All protocols involving communication were evaluated under a local area network (LAN) setting with a sustained bandwidth of 10 Gbps.

\subsection{End-to-End Evaluation on the Real Dataset}

We first evaluate the effectiveness of Agentic-PPML on ChestX-ray14 \cite{wang2017chestx}, a real-world medical imaging dataset containing over 100,000 frontal chest X-ray images annotated with 14 common thoracic disease labels. The labels are automatically mined from radiology reports using natural language processing techniques, making the dataset a widely used benchmark for automated chest pathology detection.
In our framework, we use DeepSeek-V3-0324 \cite{deepseekai2024deepseekv3technicalreport} as the general-purpose LLM to perform intent parsing and inference orchestration. The downstream task-specific model is a ResNet-50 \cite{he2016deep} classifier, deployed on the model-hosting MCP server for encrypted diagnosis.
Our system achieves a classification accuracy of 75.6\%, while maintaining strong end-to-end privacy guarantees. The total latency per secure inference is approximately 170.691 s, demonstrating that Agentic-PPML can support practical, privacy-preserving medical inference at low cost.

\subsection{Inference Overhead}

We then evaluate the efficiency of our secure inference protocol on a range of widely used convolutional neural networks (CNNs), including the Multi-Layer Perceptron (MLP), LeNet-5 \cite{726791}, AlexNet \cite{NIPS2012_c399862d}, ResNet variants \cite{he2015deepresiduallearningimage}, SqueezeNet \cite{iandola2016squeezenetalexnetlevelaccuracy50x}, and DenseNet-121 \cite{huang2018denselyconnectedconvolutionalnetworks}.
The homomorphic encryption parameters are configured for 128-bit security, with a polynomial degree of $N_p = 4096$, plaintext modulus $t = 2^{41}$, ciphertext modulus $q \approx 2^{109}$, and fractional precision $\phi = 12$.
For smaller models such as MLP and LeNet-5, we perform inference in batch mode, using batch sizes of 64 and 32, respectively. The runtime is averaged over individual samples. For larger models, we evaluate single-sample inference, and the runtime reflects the cost of processing one input end-to-end.
As shown in Table \ref{tab:sec5-available-ranges-of-k}, our privacy-preserving CNN inference protocol achieves minute-level runtime and incurs a gigabyte-level communication cost. Although the efficiency is lower than plaintext inference, the protocol achieves a practically acceptable trade-off between security and performance, making it suitable for real-world deployment where user privacy is essential.

\begin{table}
  \centering
  \caption{Performance of our secure inference protocol.}
  \label{tab:sec5-available-ranges-of-k}
  \resizebox{\linewidth}{!}{
  \begin{tabular}{ccccccc}
    \toprule
     & MLP & LeNet-5 & AlexNet  & ResNet-18 & ResNet-34 & ResNet-50 \\
    \midrule
    Runtime(s)&0.005&0.012&4.472&22.982&38.414&121.952\\
    Communication(MB)&0.296&1.028&242.219&1653.534&2748.205&8076.670\\
  \bottomrule
  \end{tabular}
  }
\end{table}


\bibliographystyle{plainnat}  
\bibliography{references}


\end{document}